\documentclass{article}

\usepackage{PRIMEarxiv}

\usepackage[utf8]{inputenc} % allow utf-8 input
\usepackage[T1]{fontenc}    % use 8-bit T1 fonts
\usepackage{hyperref}       % hyperlinks
\usepackage{url}            % simple URL typesetting
\usepackage{booktabs}       % professional-quality tables
\usepackage{amsfonts}       % blackboard math symbols
\usepackage{nicefrac}       % compact symbols for 1/2, etc.
\usepackage{microtype}      % microtypography
\usepackage{lipsum}
\usepackage{graphicx}
\usepackage{comment}
\usepackage{algorithmic}
\usepackage{algorithm,float}
\usepackage{multirow}
\usepackage{color,soul}
\graphicspath{{media/}}     % organize your images and other figures under media/ folder

\def \SysName{\textit{DisCo}}

\newcommand{\revise}[1]{{{#1}}}

\newcommand{\rerevise}[1]{{{#1}}}

\newcommand{\rererevise}[1]{{{#1}}}
\title{Optimizing DNN Compilation for Distributed Training with Joint OP and Tensor Fusion
}

\author{
  Xiaodong~Yi, Shiwei~Zhang, Chuan~Wu \\
  The Uiversity of Hong Kong \\
  \texttt{\{xdyi, swzhang, cwu\}@cs.hku.hk} \\
   \And
  Zhen~Zheng, Shiqing~Fan, Siyu~Wang, Jun~Yang, Wei~Lin \\
  Alibaba Group \\
  \texttt{\{lansong.dls, james.zz, shiqing.fsq, siyu.wsy, muzhuo.yj, weilin.lw\}@alibaba-inc.com} \\
}

\begin{document}
\maketitle

\begin{abstract}
This paper proposes \SysName, an automatic deep learning compilation module for \rerevise{data-parallel} distributed training. Unlike most deep learning compilers that focus on training or inference on a single device, \SysName~optimizes a DNN model for distributed training over multiple GPU machines. Existing single-device compilation strategies do not work well in distributed training, due mainly to communication inefficiency that they incur. \SysName~generates optimized, joint computation operator and communication tensor fusion strategies to enable highly efficient distributed training. A GNN-based simulator is built to effectively estimate per-iteration training time achieved by operator/tensor fusion candidates. A backtracking search algorithm is driven by the simulator, navigating efficiently in the large strategy space to identify good operator/tensor fusion strategies that minimize distributed training time. We compare \SysName~with existing DL fusion schemes and show that it achieves good training speed-up close to the ideal, full computation-communication overlap case.
\end{abstract}

% keywords can be removed

\section{Introduction}\label{sec:introduction}
%The complicated deployment and configuration for different types of deep learning (DL) models on diverse DL hardware have encouraged the research and development of DL compilers in recent years. 
Deep learning (DL) compilers have been studied in recent years for deep neural network (DNN) model graph optimization and training (or inference) expedition, e.g., TVM \cite{chen2018tvm}, MLIR \cite{lattner2020mlir}, Relay \cite{roesch2019relay} and XLA \cite{leary2017xla}. The DL compilers take as input the model definitions in the respective DL framework (e.g., TensorFlow \cite{abadi2016tensorflow}, MXNet \cite{chen2015mxnet}), and generate code implementation of the models on different types of DL hardware. The transformation from model definition to specific code implementation is highly optimized based on the model specification and hardware architecture, using methods including: (i) front-end optimization such as NOP elimination, zero-dim-tensor elimination, algebraic simplification, %dead code elimination, 
operator (op) fusion and layout transformation \cite{li2020deep}; and (ii) backend optimization, e.g., loop-oriented optimizations, hardware intrinsic mapping and memory latency hiding \cite{chen2018tvm}. 

Most of the existing DL compilers focus on accelerating DL model execution on a single device. In distributed training, communication among devices for parameter synchronization plays a key role in dictating the training time and resource (computation device, network bandwidth) efficiency. Compilation optimization for single-device training (e.g., op fusion) may delay inter-device communication, leading to poor computation-communication overlap and hence low distributed training efficiency (Sec.~\ref{sec:opportunities}).

Currently, only a few projects study compilation optimization in the distributed setting. %GShard \cite{lepikhin2020gshard} and Boehm {\em et al.}~\cite{boehm2018optimizing} provide distributed operator primitives in their customized compilers. 
GShard \cite{lepikhin2020gshard} extends the XLA compiler for distributed training and provides an elegant way to express a wide range of parallel computation patterns. Boehm {\em et al.}~\cite{boehm2018optimizing} use enumeration tree search with %cost-based 
structural pruning techniques for op fusion, for learning traditional machine learning (ML) models. % (not DNNs). 
However, they do not consider op fusion jointly with communication overhead in the distributed environment.

\rererevise{There are also projects focusing on model parallelism and pipeline parallelism. Megatron-LM \cite{shoeybi2019megatron} introduces an efficient intra-layer model-parallel approach to support training of very large transformer models. GPipe \cite{huang2018gpipe} and Pipedream \cite{harlap2018pipedream} propose pipeline parallelism to further improve model parallelism, by pipelining forward computation and backward propagation across several micro-batches. CoCoNet \cite{jangda2022breaking} enables optimization of data-, model- and pipeline-parallel workloads in large language models by introducing a domain-specific language that easily expresses distributed training of models.}

This paper focuses on front-end compilation optimization to expedite synchronous data-parallel training. Op fusion strategies have been studied as one of the most important optimization methods to reduce computation overhead \cite{leary2017xla,RAMMER,zheng2020ansor}. Tensor fusion has been shown to play an important role in reducing the communication overhead \cite{bao2020preemptive,peng2019generic,sergeev2018horovod}. We inspect the performance trade-off caused by op fusion and tensor fusion in distributed training, and advocate joint op and tensor fusion optimization.
We propose \SysName, an automatic module to jointly optimize computation and communication fusion over %the whole High Level Optimizer (HLO) module of 
a whole distributed DNN training graph. Existing rule-based op fusion strategies %require significant engineering efforts and 
rely heavily on expert experience, and are often less than optimal due to limited exploration of the solution space. % of pruning large size of solution space with their heuristics. 
\SysName{} adopts a search-based algorithm to identify optimized joint fusion strategies. %using an effectively-built simulator.
We summarize main contributions of \SysName{} in the following:

%On the other hand, in distributed training, the influence of communication cannot be ignored when applying fusion optimization: a good fusion strategy in a single device may delay the gradient synchronization among devices in distributed training.

%We advocate jointly optimization of computation operator fusion and communication tensor fusion,

$\triangleright$ We propose an automatic compilation module to jointly optimize op and tensor fusion for distributed training of DNN models, that expedites computation and communication separately while maximally overlapping their execution.

$\triangleright$ Op fusion and tensor fusion, two conventionally separated optimization passes, are unified into a joint strategy space. A %cost-based
backtracking search algorithm is designed to efficient prune the large strategy space to identify op/tensor fusion solutions that jointly minimize distribution DNN training time.

$\triangleright$ A {\em Fused Op Estimator} is built based on a graph neural network (GNN) model to predict the execution time of fused ops. An efficient simulator is created to estimate the end-to-end execution time of a distributed DNN training graph using the Fused Op Estimator, and serves as a cost model to our search algorithm. 

$\triangleright$ We implement \SysName~based on JAX \cite{jax2018github}, an XLA-based framework for generating high-performance accelerator code %from pure Python and Numpy ML programs, and the optimized code generation is 
in a manner completely transparent to DNN model developers. To use \SysName, a developer only needs to specify two environment variables, not changing a single line of their model code. \rerevise{\SysName{} is open-sourced at https://github.com/TPDS-Submission/Disco.}%We plan to open source \SysName~to the community.

$\triangleright$ We carry out extensive experiments training state-of-the-art DNN models in GPU clusters, and carefully compare \SysName~with existing DL fusion schemes. \SysName~achieves up-to 26.73\% training acceleration, close to the maximal speed-up achievable with ideal, full computation-communication overlap. Interestingly, we observe that our joint op and tensor fusion optimization not only increases communication-computation overlap, but also reduces computation time and communication time, separately, as compared to representative single-device fusion designs.

\section{Background and Motivation}
\label{sec:motivation}

\subsection{Deep Learning Compilation}
\begin{comment}
Existing DL libraries are usually customized for specific types of DL hardware. For example, Basic Linear Algebra Subprograms (BLAS) libraries (e.g. cuBLAS \cite{nvidia2008cublas}) are highly optimized and served as the basics for efficient computation of DL models on GPUs. Hardware vendors have also proposed specially optimized libraries for DL computation (e.g. cuDNN \cite{chetlur2014cudnn}). % including forward and backward convolution, pooling, normalization, and activation. However, the drawback of relying on the libraries is that they usually fall behind the rapid development of DL models and new types of DL hardware.
\end{comment}

%To address the drawback of DL libraries and tools, as well as 
To alleviate the dependence on customized DL libraries and the burden of manually optimizing
DL models on each type of hardware, domain
specific DL compilers have been built \cite{chen2018tvm}\cite{vasilache2018tensor}\cite{rotem2018glow}\cite{leary2017xla}. %, e.g., TVM \cite{chen2018tvm}, Tensor Comprehension \cite{vasilache2018tensor}, Glow \cite{rotem2018glow} and XLA \cite{leary2017xla}.
% The DL compilers take the model definition described in the DL frameworks as input, and generate efficient code implementation on various DL hardware. 
DL compilers incorporate DL-oriented optimizations such as layer and op fusion, to generate highly efficient code for training or inference. %In addition, existing DL compilers
%They may also leverage mature tool-chains from general-purpose compilers (e.g., LLVM \cite{lattner2004llvm}) to achieve better portability across diverse hardware architectures. 
Similar to traditional compilers, DL compilers utilize {\em intermediate representation} (IR) as the abstraction of a DNN model for optimization, including high-level IR which represents the control flow and the dependency among the operators and the data, and low-level IR which reflects hardware characteristics such as memory allocation. DL compilers adopt the layered design, including the front-end optimization (transforming the DNN model into the high-level IR and performing graph-level optimization such as dead code elimination and op fusion) and the back-end optimization (transforming the high-level IR into low-level IR and performing hardware-specific optimization). %such as memory allocation and loop fusion).
Our study focuses on high-level IR optimization at the DNN graph level.

\begin{figure}
  \centering
  \includegraphics[width=0.5\columnwidth]{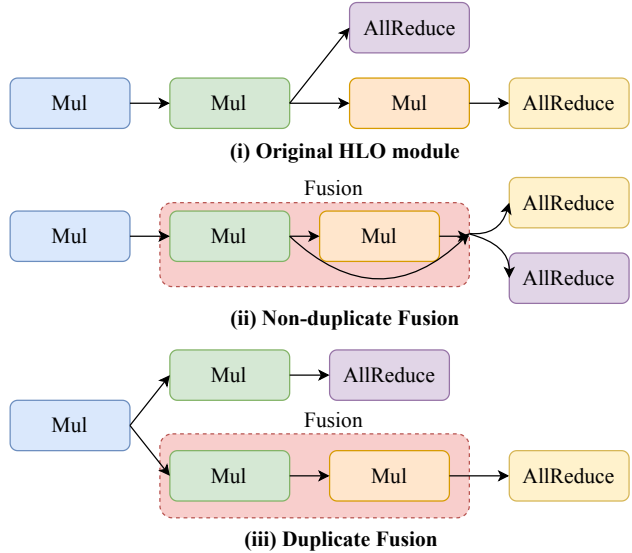}
  \caption{Non-duplicate fusion and duplicate fusion. An arrow represents gradient/activation passing.
  }
  \label{fig:duplicate}
\end{figure}

\subsection{Computation Operator Fusion}
Op fusion \cite{chen2018tvm}\cite{boehm2018optimizing}\cite{li2020deep} is a graph-level optimization that combines multiple computation operators into a single kernel without storing the intermediate results in device memory (e.g., global memory on a GPU). It enables better utilization of computation devices, eliminates device memory allocations for intermediate results, and reduces kernel launch and synchronization overhead, leading to substantially reduced model execution time. 
Op fusion has been enabled in a number of DL libraries such as TensorFlow XLA \cite{leary2017xla}, Intel Nervana Graph \cite{cyphers2018intel} and TVM \cite{chen2018tvm}. 

To carry out op fusion, typically an op is selected, and then one of its predecessor ops (whose output this op consumes) is chosen to fuse with this successor op. If the chosen predecessor op has multiple successor ops, two main fusion approaches exist: non-duplicate fusion and duplicate fusion \cite{li2020deep} (exemplified in Fig.~\ref{fig:duplicate}). With non-duplicate fusion, the predecessor op is directly fused into the successor op; the output of the predecessor (e.g., gradients) is available for other ops only after the completion of the fused op. %, and then can be consumed by other operators. 
With duplicate fusion, the predecessor op is not only fused into the successor op, but also recomputed outside the fused op (so that its output can become available earlier). Op fusion can be carried in a recursive manner over the entire DNN graph: a fused op can be further fused with its predecessor or successor, using a duplicate or non-duplicate fusion approach. 

\begin{figure}[t]
  \centering
  \includegraphics[width=0.5\columnwidth]{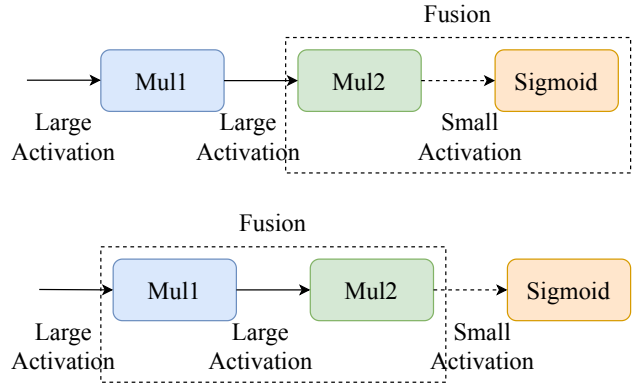}
  \caption{A case in RNNLM: the order of ops to consider for fusion influences performance significantly.}
  \label{fig:fusion_motivation}
\end{figure}

The order of ops to consider for fusion is typically determined by heuristics or learning-based methods \cite{abdolrashidi2019learning}\cite{long2018fusionstitching}. 
For example, in XLA, ops are chosen according to a pre-defined post order, and any device memory and computation savings due to fusing the op with a selected predecessor op are evaluated. 
Such op order-based fusion may not be effective as earlier fusion of some ops may prevent better fusion opportunities for ops considered later. 
Consider a case of fusing two ops in RNNLM \cite{ji2016latent} in Fig.~\ref{fig:fusion_motivation}, 
where an element-wise multiplication op (Mul1) produces large activations to another multiplication op (Mul2) and Mul2 produces small activations to a Sigmoid function. If Sigmoid ranks higher in the op ordering and is fused with Mul2,   
the performance does not improve much, since the size of intermediate data (activations) transferred between on-chip memory (local memory for the execution thread) and device memory does not change significantly. 
If Mul1 and Mul2 are fused instead, activations produced by Mul1 remain in on-chip memory for Mul2, substantially reducing data transfer to/from device memory. %Therefore, after Reduce returns the result, the amount of transferred data has decreased dramatically. Thus, as it can be seen, there should be more priority to fuse the more appropriate nodes. An inefﬁcient fusion algorithm may not improve the performance much and sometimes, might even hurt the performance. 

\begin{comment}
Most op fusion systems use heuristics or manual declaration for op fusion. %, and focus on single-device DNN graph optimization. 
Fusionstitching \cite{long2018fusionstitching} fuses ops with critical-path reduction based heuristics. % and considers both producer-consumer fusion opportunities and also fusion of fine-grained ops in the same DNN layer. 
Google team \cite{abdolrashidi2019learning} propose a learning-based framework to decide the order of op fusion. 
\end{comment}

Besides, majority of the existing op fusion systems focus on single-device DNN graph optimization \cite{chen2018tvm,jia2019taso,roesch2019relay,cyphers2018intel}. %%\cwu{add citations}. %with poor support for distributed training.
%It is challenging to navigate through the complete op fusion solution space to identify the best fusion strategies for a DNN graph.

\subsection{Communication Tensor Fusion}

In distributed training, data parallelism has been most widely adopted in practice. The training dataset is partitioned into mini-batches at each device. In each training iteration, each worker (device) maintains a replica of the DNN model and carries out Forward Propagation (FP) and Backward Propagation (BP) computation on a mini-batch; gradients from different devices are aggregated before being applied to update model parameters. We focus on accelerating data-parallel training. % in this paper.

%In each iteration of distributed DNN training, gradient aggregation among workers is necessary for synchronization. 

AllReduce is a collective instruction, popularly used for parameter synchronization in data-parallel training. It sums (or averages) the gradients from all devices using a ring or tree based algorithm \cite{jeaugey2017nccl}, and disperses the aggregated gradients to the devices for parameter update  \cite{patarasuk2009bandwidth}. Commonly one AllReduce instruction is carried out for each gradient tensor produced; the default sizes of tensors in existing DL frameworks (e.g., TensorFlow, PyTorch) may not be ideal for efficient bandwidth utilization. There are usually a large number of small tensors (e.g., over 50\% communication tensors in ResNet50 \cite{he2016deep} and Transformer \cite{dai2019transformer} are less than 1MB in size \cite{bao2020preemptive}). Such small tensors incur large communication overhead in relation to the short transmission time, e.g., time spent on negotiation/synchronization among workers before actual gradient transfer, %and low bandwidth utilization.  %can not fully utilize network bandwidth during communication, due to the relatively larger portion of overhead 
 %The overhead is especially non-trivial for AllReduce instructions, due to 
 which is especially substantial in view of the strict synchronization among workers required by AllReduce. %Existing literature \cite{bao2020preemptive} observe that over 50\% communication tensors in ResNet50 \cite{he2016deep} and Transformer \cite{dai2019transformer} are less than 1MB in size. They also observe that when the tensor size is small (less than 1MB), the communication time does not decrease linearly with the decrease of tensor size, because the communication overhead (time for message transmission, negotiation time among all workers before performing all-reduce on a tensor) is too large to ignore as compared to the short transmission time. For DNNs with large amount of tiny gradient tensors, the overhead in AllReduce communication can be significant.

Tensor fusion advocates fusing multiple small gradient tensors together before executing the AllReduce instruction on the fused tensor. The size of the fused tensor is the sum of sizes of the small tensors. %but the fused all-reduce instruction is only ready for execution when all the small all-reduce instructions are ready. 
Tensor fusion leads to better bandwidth utilization (with less communication overhead relative to actual gradient transfer); however, start time of the fused AllReduce is delayed, leading to a trade-off effect on the training time. % between bandwidth saving and the delay of the transmission.

\subsection{Opportunities}
\label{sec:opportunities}

To accelerate data-parallel training, %of DNN models on large datasets, 
existing proposals improve the computation graph on each device with single-device compilation optimizations (e.g., op fusion) 
and optimize inter-device communication, separately \cite{jax2018github,leary2017xla}. Op fusion effective on individual devices may be non-optimal or even bring no benefit in distributed training. Op fusion typically merges as many ops as possible to reduce device memory usage and kernel launches; the output of those ops may only be available after the fused op is completely executed. For example, gradients produced by backward propagation ops need to be transferred to other devices for aggregation; fusion of BP ops may delay gradient communication, leading to less computation-communication overlap and hence longer resource idling time.

\begin{figure}[!t]
  \centering
  \includegraphics[width=0.75\columnwidth]{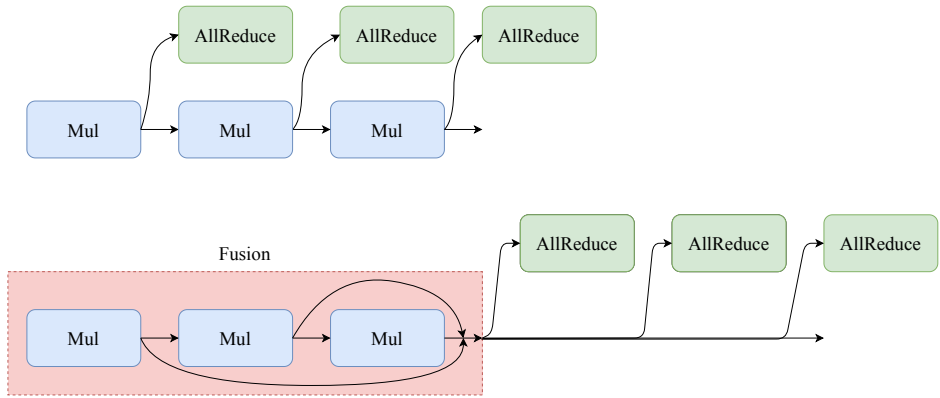}
  \caption{%A case showing 
  Delayed communication due to op fusion.}
  \label{fig:jointly_motivation}
\end{figure}

Fig.~\ref{fig:jointly_motivation} gives an example. Suppose %a fusion strategy fuses
the three \texttt{Mul} ops are fused, such that AllReduce instructions of gradients produced by the 3 ops are delayed until after the fused op is done. If the delay exceeds computation time reduction, such op fusion increases the training time.

Only a few systems enable DL compilation in distributed training. %TensorFlow XLA \cite{leary2017xla} implements AllReduce; 
Based on AllReduce primitives provided by XLA \cite{leary2017xla}, JAX \cite{jax2018github} %improves the generation of the High Level Optimizer (HLO) module (a high-level IR defined in XLA) by 
groups the whole processing logic of a DNN model, including the AllReduce instructions, into a single High Level Optimizer (HLO) module (a high-level IR defined in XLA). JAX currently only supports multi-node training across TPU servers rather than GPU servers, and uses rule-based heuristics for op fusion. Rule-based op fusion highly depends on expert experience, and the rule suitable for one model may not fit other models \cite{li2020deep}.
Further, its computation optimization is separated from communication optimization: op fusion is first conducted, and AllReduce combiner optimization (combining multiple AllReduce instructions together based on a pre-defined tensor size threshold) is performed after op fusion optimization is done. %This leads to the drawback as analyzed above.  %Community
It has been reported \cite{issue2,issue1} that directly applying XLA in distributed training may prolong per-iteration training time (20\% slower when training a transformer-based NMT model with Adam Optimizer \cite{issue2}), as compared to not applying XLA, since communication can be seriously delayed.

%This process has multiple drawbacks: (1) rule-based op fusion optimization highly depends on expert experience, and the rule suitable for one model may not fit other models; 
%(2) the computation optimization may affect the communication during the back propagation: when fusing multiple operations together, the outputs can only be fetched after the execution of the fused operation. If there are some outputs needs to be transferred to other devices(e.g. the gradients needs to be aggregated across all devices), the operation fusion might delay the transmission. So, separately optimizing computation operation fusion and communication all-reduce combination is far away from the optimal.Fig.~\ref{fig:jointly_motivation} illustrates this situation: The \texttt{Conv2D\_BP} instructions computes the gradients of the Conv2D layers. In single-device training, the fusion strategy fuses these 3 instructions into a single fused instruction to eliminate the frequent data transfer from on-chip memory and device memory, which can reduce the overall computation time. But in distributed training, The outputs (gradients) of the 3 instructions need to be aggregated across all nodes using AllReduce. The strategy used in single-device training delays the transmission of the AllReduce instruction. If the delay exceeds the benefit from computation time reduction, the fusion makes the overall computation time even longer.

There is a trade-off between computation efficiency and communication channel utilization in distributed training, when both op fusion and tensor fusion are adopted. %\revise{As one of the most important optimization methods to reduce computation overhead, op fusion strategies need to be more carefully designed in a distributed environment, where tensor fusion may play an important role in reducing the communication overhead. We explore these two aspects of training optimization in the distributed setting in this paper. }
We advocate a search algorithm to jointly optimize op and tensor fusion, striking a good balance between computation and communication efficiencies and achieving overall training acceleration. %not only reduce the computation and communication time separately, but also maximize their overlap.

\subsection{Challenges}
\label{sec:challenges}

Exploring the opportunities comes with challenges.

\noindent\textbf{Large search space for joint fusion.} A DNN model usually consists of thousands of computation and communication instructions, resulting in a huge search space with various op/tensor fusion combinations. Naive enumeration of possible solutions without pruning is infeasible. We design an efficient backtracking algorithm to prune the search space effectively.

\noindent\textbf{Time- and resource-consuming to evaluate search candidates in real environments.} Unlike single-device training, evaluating each possible solution produced by the search algorithm by running the modified DNN model in a real distributed environment is time- and resource-prohibitive. %It requires all the devices carry out trial runs of thousands of possibles solutions until the search algorithm finds a good strategy. %To solve this problem, 
We build an efficient simulator to estimate the execution time of possible strategies produced by the search algorithm, eliminating the need of heavy real-world trial runs.

\noindent\textbf{Difficulty in accurate execution time prediction of fused ops.} %Using a simulator to predict the execution time of a DNN model is a common way to save time and computation resource. 
Simulators have commonly been used to predict DNN training time under different device placements \cite{jia2019taso,jia2018beyond} or with different execution scheduling strategies \cite{yi2020fast,yi2020optimizing}, based on profiled execution time of individual ops. In our case, fused ops that have never been seen before may well be produced. Execution time estimation for fused ops is not easy: even if execution time of each original op is profiled, the interaction among these ops is unknown, and cannot be profiled unless we implement every unseen fused op. %is impossible to be profiled before meeting the unseen fused instruction. It is hard to estimate the per-iteration training time of the DNN model without an accurate estimation of these fused instructions. 
Further, execution time of fused op is tightly related to the architecture of the processor, as well as the back-end optimization applied during compilation such as loop fusion, tiling and loop unrolling \cite{li2020deep}.
Architectural features and compiler code generation interact in extremely complex ways \cite{chen2018tvm,leary2017xla,RAMMER}. It is very hard to build a white-box analytical model describing details of the processor or effects of all compiler passes, and their interactions. %Designing a white-box time estimator for these fused instructions by considering all dimensions mentioned above is extremely time consuming and error-prone.

%\vspace{1mm}
We design a GNN-based model for execution time prediction of fused ops. GNNs have been adopted and achieved satisfying performance for various graph-based learning purposes, %e.g., node classification
\cite{zhou2020privacy}\cite{xu2018powerful}
%, link prediction \cite{
\cite{zhang2018link}\cite{kazemi2018simple}
%}and recommendation
\cite{fan2019graph}\cite{wu2019session}. It takes as input graph-structured data and learns the structural information based on graph connectivity and node/edge features. We exploit a GNN to learn the execution time from the op fusion graph. \revise{We focus on optimizations that preserve model accuracy (exactly the same before and after optimization), and hence do not consider staleness options which may compute gradients based on the last round of weights while communicating the gradients of this round} \cite{harlap2018pipedream}.
%Our GNN model takes as input inter-connectivity and features of ops to fuse, and predicts the execution time of fused op.
\section{System Design}
\subsection{\SysName~Overview}
\label{sec:overview}

\begin{figure}[t]
  \centering
  \includegraphics[width=0.75\columnwidth]{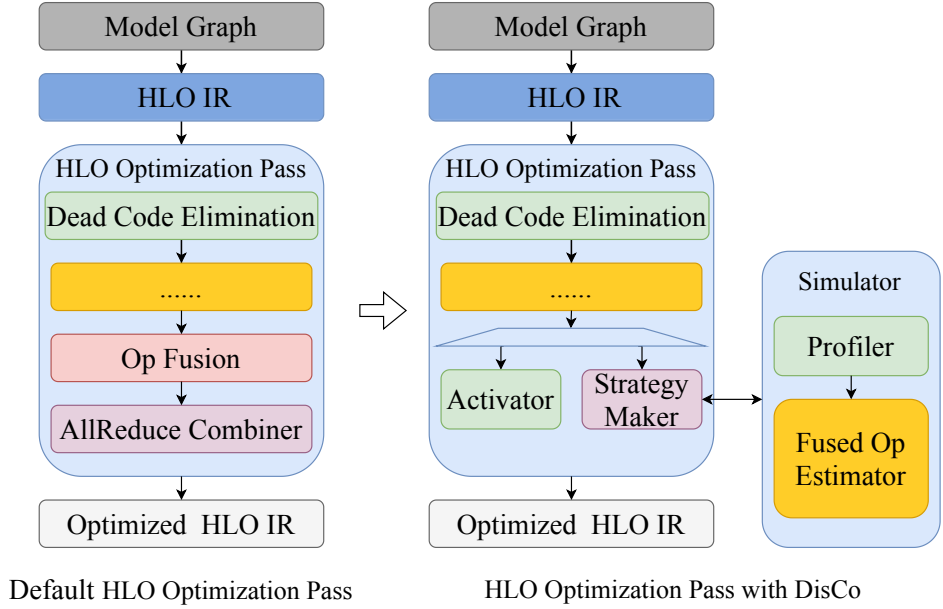}
  \caption{Overall architecture of \SysName.} 
  \label{fig:model}
\end{figure}

\SysName~is designed as an %customized 
optimizer for TensorFlow XLA's %High Level Optimizer 
HLO IR, to produce optimized fusion strategies for both computation operators and communication tensors. \SysName~takes as input the HLO module of a whole DNN model, and produces an optimized HLO module for further back-end compilation optimization. Fig.~\ref{fig:model} shows the overall architecture of \SysName.

\SysName~has two phases: {\em Search Phase} and {\em Enactment Phase}. In the {\em Search Phase}, the {\em Strategy Maker} (which runs on the master node in the TensorFlow framework) uses a backtracking algorithm to jointly search for the best op/tensor fusion strategies for distributed DNN training. In the {\em Enactment Phase}, the {\em Activator} (residing on each worker) retrieves the HLO module optimized with the best strategies to each worker, and activates the strategies for distributed training.

To facilitate the backtracking search algorithm in {\em Strategy Marker}, the {\em Simulator} estimates the per-iteration training time of the DNN model using candidate strategies that the backtracking algorithm generates. A GNN-based {\em Fused Op Estimator} predicts the execution time of fusion ops, to serve the {\em Simulator}. The {\em Profiler} runs the DNN model to record execution time of individual ops and prepares the training data for the GNN model of {\em Fused Op Estimator}.

\SysName~provides a simple switch for developers to alter the phase of the system: when setting an environment variable \texttt{ENABLE\_SEARCH} to 1, the search phase is activated and backtracking search is used to identify the best fusion strategies; when \texttt{ENABLE\_SEARCH} is 0, %the system activates the normal execution 
enactment phase starts and distributed training is activated using the best strategies found in the search phase.

\subsection{Strategy Maker}
\label{sec:strategymaker}

Our strategy space includes combinations of the following set of strategies:
(i) fusion strategy for each computation op: no fusion, or fusing the op with a predecessor op $p$, $\forall p$ among the op's predecessor ops in the current HLO (which can be original op or fused op); %whether the op should be fused with one of its predecessor ops and which predecessor.
(ii) fusion approach for a predecessor op which has multiple successors: %and it needs to be fused to one of its outputs, 
 duplicate fusion or non-duplicate fusion (Fig.~\ref{fig:duplicate});
(iii) fusion strategy for each AllReduce %gradient tensor
instruction: no fusion, or combining the tensor with any of the neighboring original or fused gradient tensors. A neighbor gradient tensor is produced by a BP gradient computation op that is a successor or a predecessor to the op producing the current gradient tensor.

The goal is to minimize per-iteration training time of the DNN model, i.e., end-to-end execution time of the HLO module in the distributed setting (including execution time of computation ops and AllReduce instructions). %both computation and communication time
 The {\em Strategy Maker} exploits a backtracking algorithm to explore the joint strategy space %the three dimensions mentioned above 
 and exploits the {\em Simulator} to guide the search directions.

\section{Strategy Framework}
\label{nn_design}
\subsection{Activator}
%The {\em Activator} component resides on both the master node in TensorFlow framework (which creates the entire graph, and initialize nccl communication channel) and every worker.
In the {\em Enactment Phase}, %the {\em activator} is responsible for activating the best strategy in all nodes.
the {\em Activator} on the master node fetches the optimized HLO module generated by {\em Strategy Maker}, and broadcasts it to each of the other workers. The activators in other workers %waits for the broadcasting of 
receive the HLO module %from the master node. %After the broadcasting, All workers 
and then execute the optimized HLO module together, i.e., carry out distributed training using the optimized strategies.

\subsection{Simulator}
In the {\em Search Phase}, the {\em Simulator} is used as a cost model to drive the backtracking algorithm in the {\em Strategy Maker}. It simulates training according to the strategies produced by the {\em Strategy Maker}, and estimates the per-iteration training time using profiled data from the {\em Profiler} for individual ops and the {\em Fused Op Estimator} for fused ops.
%Two auxiliary modules are used for building the {\em Simulator}:

\vspace{1mm}
\noindent\textbf{Profiler}. It profiles distributed training of the given DNN model to obtain execution time of each HLO instruction and communication time of each AllReduce instruction across different devices. The execution time of each HLO instruction is recorded and indexed by its \texttt{op\_code} and input shape. %For AllReduce instructions on tensors of different sizes, 
We build a linear regression model for communication time prediction of AllReduce instructions according to the tensor size: $T=Cx+D$, where T is the predicted execution time of the AllReduce instruction, x is the size of the gradient tensor, C reflects the bandwidth and D is the communication overhead in AllReduce instructions. Normally, AllReduce execution time is affected by multiple factors including tensor size, network topology and bandwidth, and the communication library in use. In our scenario, the time is most relevant to the tensor size as other factors are fixed. Taking ring AllReduce as an example, \revise{if the NICs work at the full-duplex mode,} the communication time can be computed as $T=\frac{2(N-1)x}{B\times N}$ \cite{kim2019parallax}, where N is the number of devices %, x is the tensor size
and B is the smallest bandwidth between any device pair along the ring; T is linear with x when B and N are fixed, ensuring a simple linear regression model is accurate enough for our prediction purpose.
\begin{figure}[]
  \centering
  \includegraphics[width=0.75\columnwidth]{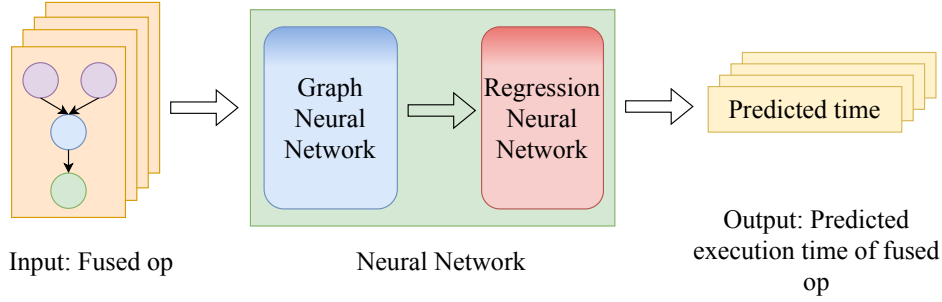}
  \caption{%The training framework of 
  GNN-based Fused Op Estimator.}
  \label{fig:nn}
\end{figure}

\vspace{1mm}
\noindent\textbf{Fused Op Estimator}. %Fused ops %are not shown before, their computation time cannot be estimated by the regression model built based on
%are not profiled by the {\em Profiler}. 
We design a GNN model to predict execution time of each fused op, which %takes as input the single fused computation in the hlo module and predict the execution time of the fused computation.
takes as input inter-connectivity and features of ops to fuse (i.e., execution time of individual ops), and predicts execution time of the fused op.

\subsection{GNN-based Fused Op Estimator} 
\label{sec:GNNdesign}

Since each fused op consists of multiple original ops, a fused op can be regarded as a subgraph of the DNN model graph, whose nodes are the original ops and edges are the dependencies among them. A GNN is a nice fit for learning features of the subgraphs for fused op execution time prediction\rerevise{: the GNN takes the op type, input and output sizes, execution time of each original op and the data dependency among them as input features, and learns the execution time of the fused op as output; our prediction problem can be regarded as a GNN graph classification and regression job, which has been studied in the literature \cite{errica2019fair}\cite{cai2018simple}\cite{li2021ood}.} Based on the GNN, the {\em Simulator} can further calculate the execution time of the whole HLO module. We do not use a GNN to directly predict the execution time of the whole HLO module, since the computation time of individual ops can be profiled and the communication time of AllReduce instructions can be estimated using the linear regression model, and we can have more accurate estimation accordingly. %there is no need for GNN to predict the per-iteration training time of the whole HLO module, which requires much larger GNN structure and much more training datasets compared with predict the execution of a single fused op.  
An illustration of our GNN model is in Fig.~\ref{fig:nn}.

\begin{comment}
\begin{figure}
  \centering
  \includegraphics[width=\columnwidth]{figure/Embedding.eps}
  \caption{A graph neural network transforms the raw information on each original op into a vector representation. This example shows the steps of message passing and summarization.}
  \label{fig:embedding}
\end{figure}
\end{comment}
\subsubsection{Feature Encoding}

%Since the original ops in different fused ops have different number and control dependency, 
The GNN layers create a flat feature vector for each fused op %by treating the structure of a fused op (including ops to fuse together and their dependencies) as a graph, 
by encoding its subgraph into a set of embeddings. %Fig.~\ref{fig:embedding} shows an example of original node embedding and fused op embedding.

\vspace{1mm}
\noindent{\em Original op embeddings.} The GNN takes as input the %graph representation of the fused op, in the form of
following subgraph information: (1) an op feature matrix, where each row corresponds to one original op and contains the op's attributes, including execution time, input and output sizes, op type (e.g. Conv2D, MatMal); (2) an adjacency matrix describing data dependencies among the ops. 
It generates a per-node embedding vector $e_i$, by encoding attributes of immediate neighbors of op $i$ using multi-head attention layers \cite{vaswani2017attention}:

%\vspace{-6mm}
\begin{equation}
    \mathbf{e}_i = \Vert_{k=1}^{K}\sigma(\sum\limits_{j\in\mathcal{N}_i}\gamma_{ij}^{k}W^{k}\mathbf{e}^{'}_j)\nonumber
\end{equation}
\vspace{-3mm}

\noindent Here $K$ is the number of heads of the multi-head attention layer, $\Vert$ denotes concatenation of the output of each head, $\sigma$ is a non-linear transformation, ${N}_i$ is the set of neighbors of op $i$ including $i$ itself, $\gamma_{ij}$ is the correlation coefficient between feature vectors of op $i$ and op $j$, $W$ is the weight vector to be learned, and $\mathbf{e}^{'}_j$ is the output embedding of op $j$ from the previous attention layer.

\vspace{1mm}
\noindent{\em Fused op embeddings.} A fused op embedding $\mathbf{y}$ is generated by encoding information from all original ops in the fused op:

\vspace{-3mm}
\begin{equation}
    \mathbf{y} = \sigma(\sum\limits_{i\in\mathcal{N}}W\mathbf{e}_i)\nonumber
\end{equation}
\vspace{-3mm}

\noindent where $\mathcal{N}$ contains all the original ops in the fused op. 

%\begin{comment}
\subsubsection{Regression Neural Network}
The fused op embeddings are fed into a regression neural network \cite{kaufman2020learned} for execution time prediction, which consists of a number of Fully Connected (FC) layers followed by a Relu activation function. %The output of the network is a scalar value representing the execution time of the fused op.
%\end{comment}

\subsubsection{Model Training}
\label{sec:alg}
The graph neural network and regression neural network are trained together in a supervised manner, %In each round, 
using a sampled set of fused ops, $G$ (see Sec.~\ref{sec:simul_imp} for details of producing GNN training samples). % are sampled as input to the GNN. 
 For each fused op, predicted execution time is produced by the GNN model. The objective is to minimize the overall loss over the $|G|$ fused ops:
 
%\vspace{-6mm}
\begin{equation}
L(\theta) =\frac{1}{|G|}\sum\limits_{g\in G}\log(\mathbf{y}_g-\mathbf{y}^{'}_g)^2
\end{equation}
\vspace{-3mm}

\noindent where $\theta$ is the set of weights in the GNN model to learn, and $\mathbf{y}_g$ and $\mathbf{y}^{'}_g$ are the predicted and profiled execution time of fused op $g$, respectively. We use Adam Optimizer \cite{zhang2018improved} to minimize the loss function.

\subsection{End-to-end HLO execution time estimation}
The simulator computes end-to-end execution time of an HLO module, by simulating \rerevise{scheduling} process of the HLO on \rerevise{one} device and taking AllReduce communication among this device and others into account.
\rerevise{The complete scheduling process %of the entire simulation 
can be described as follows.} A ready queue is maintained, consisting of computation ops whose dependencies have been cleared. Iteratively, a ready op is removed from the head of the queue, \rerevise{and the completion time of the op is computed according to the completion times of its predecessors and its own execution time. Then, this op's successors can be appended to the tail of the queue if the respective successor's dependencies are all cleared. }
\revise{AllReduce instructions are executed in order of production of their respective gradient tensors (which can be original tensor or fused tensor). An AllReduce instruction starts after its gradient tensor is produced (in case of a fused tensor, after all tensors composing the fused tensor have been produced) and the communication channel becomes clear, and its execution can overlap with the execution of computation ops in time.} 
%The duration of an AllReduce instruction is estimated using the linear regression transfer time model. The end-to-end execution time of the HLO module is then computed as the latter of the end time of execution of all ops and AllReduce instructions.
The simulator serves as a cost model $Cost(H)$, where H indicates the candidate HLO module, in our strategy search algorithm.

\subsection{%Cost Based 
Backtracking Search}
\label{sec:searchalg}

\begin{algorithm}[t]
    \caption{Backtracking Search}
    \label{alg:search}
    \begin{algorithmic}[1]
    \STATE{ {\bf Input:} input HLO module $\mathcal{H}_{in}$, optimization method set $\mathcal{S}$, cost model Cost(·), parameters $\alpha$ and $\beta$. }
    \STATE{{\bf Output:} {optimized HLO module.} }

    \STATE{ $\mathcal{Q} := \{\mathcal{H}_{in}$} \} \# $\mathcal{Q}$ is a priority queue sorted by $Cost(\cdot)$.
    \STATE{ $unchanged\_counter := 0$} \# a counter to record the number of steps in which $\mathcal{H}_{Opt}$ has not been changed.
    \WHILE{$\mathcal{Q} \neq \{\}$ and $unchanged\_counter < 1000$}
        \STATE{$\mathcal{H} := \mathcal{Q}.dequeue()$}
        \FOR {optimization method $s \in \mathcal{S}$} 
            \STATE{\# Generate a random value ranging from 0 to $\beta$.}
            \STATE {$n := Random(0,\beta)$} \\
            \STATE{\# %$RandomApply(\mathcal{H},s,n)$ 
            randomly apply s on $\mathcal{H}$ for $n$ times.}
            \STATE {$\mathcal{H}^{'} := RandomApply(\mathcal{H},s,n)$ }
            \IF{$\mathcal{H}^{'}$ is valid}
                \IF{$Cost(\mathcal{H}^{'}) < Cost(\mathcal{H}_{Opt})$}
                    \STATE{$\mathcal{H}_{Opt} := \mathcal{H}^{'}$}
                    \STATE{$unchanged\_counter=0$}
                \ELSE
                    \STATE{$unchanged\_counter+=1$}
                \ENDIF
                \IF{$Cost(\mathcal{H}^{'}) \le \alpha \times Cost(\mathcal{H}_{Opt})$}
                    \STATE{$\mathcal{Q}.enqueue(\mathcal{H}^{'})$}
                \ENDIF
            \ENDIF
        \ENDFOR
    \ENDWHILE
    \RETURN $\mathcal{H}_{Opt}$
    \end{algorithmic}
\end{algorithm}

\rerevise{The strategy maker exploits a backtracking search algorithm to explore
the joint strategy space.} Alg.~\ref{alg:search} summarizes our search algorithm. Corresponding to the three types of strategies (Sec.~\ref{sec:strategymaker}), three optimization methods ($\mathcal{S}$) are explored in our search:

(i) Randomly choose one computation op, and fuse it with a randomly chosen predecessor op in the current HLO module; if the selected predecessor op $p$ has multiple successor ops, redirect the output of the fused op to $p$'s other successors.  (Fig.~\ref{fig:duplicate}(ii)).

(ii) Randomly choose one computation op, and fuse it with a randomly chosen predecessor op in the current HLO module; if the selected predecessor op $p$ has multiple successor ops, duplicate $p$ and direct the output of the replica to other successors of $p$ (Fig.~\ref{fig:duplicate}(iii)).

(iii) Randomly choose one AllReduce instruction, and combine it with a randomly chosen neighbor AllReduce instruction. A neighbor AllReduce instruction corresponds to a (fused) gradient tensor produced by a (fused) BP gradient computation op, neighbor to the op producing the chosen tensor. 

The reasons of potentially duplicating a predecessor op (as in (ii) above) are as follows: on one hand, the time needed for re-computing the op could be smaller than data transferring time between on-chip memory and device memory when not using fusion, if the size of activations produced by the op is large; on the other hand, the output of the duplicated op can be transferred immediately to other successor ops, without waiting for completion of the fused op. %In our design, we consider all ops, including %heavy communication ops (i.e., AllReduce instructions that transfer large tensors and computation ops, because our op fusion strategy is not intended for heavy communication ops only, but can greatly reduce execution time of small computation ops by fusing them together.

To explore the strategy space for producing an optimized HLO module, a priority queue $\mathcal{Q}$ is maintained for backtracking: some candidate HLO modules, produced during the search process, are buffered in order of their $Cost()$ (i.e., end-to-end execution time) for further optimization; the optimization methods are recursively applied to these candidate HLO modules. %fused ops can be further fused with their successors.
Initially, the original HLO module is enqueued into $\mathcal{Q}$. In each search step, the algorithm dequeues the HLO module $\mathcal{H}$ from the head of $\mathcal{Q}$. Each of the three optimization methods is applied on $\mathcal{H}$ for $n$ times \rerevise{(noted as $RandomApply$ in Alg.~\ref{alg:search})}, where $n$ is a random number within the range of $0$ and $\beta$ ($\beta$ is a positive integer). If the obtained new HLO module $\mathcal{H}^{'}$ is valid (i.e., not including op fusion which should not be done, e.g., the op is a parameter type or control-flow op such as \texttt{switch} and \texttt{while}), %functionally equivalent to $H$, 
we compare the execution time of $\mathcal{H}^{'}$ with that of the best HLO module,$\mathcal{H}_{Opt}$, identified so far, and record $\mathcal{H}^{'}$ as the best HLO module if its cost is smaller. On the other hand, if $\mathcal{H}^{'}$'s execution time is no larger than $\alpha$ ($\alpha \ge 1$) times that of $\mathcal{H}_{Opt}$'s, it will be enqueued into $\mathcal{Q}$ for backtracking and optimization again in further steps. The search process continues until $\mathcal{Q}$ is empty or $\mathcal{H}_{Opt}$ remains unchanged for a number of steps (1000), and returns the best HLO module $\mathcal{H}_{Opt}$ identified.

\rerevise{$\alpha$ and $\beta$ are two key hyper-parameters in our backtracking search algorithm. $\beta$ determines the probability to fuse more ops in one step. As described in Alg.~\ref{alg:search}, we} evaluate the \rerevise{modified} HLO module (using the simulator) once every n times rather than each time after applying an optimization method, \rerevise{where $n$ is randomly generated for applying each optimization in each step with the upper bound $\beta$. This is} because the change of the HLO module is subtle after only applying an optimization method once, %making the simulator hard to differentiate the HLO module before and after modification; second, 
as well as to reduce the evaluation time for expedited search. %evaluating the modification HLO module every n times after applying the optimization can reduce the evaluating rounds and significantly accelerate the search process.
\rerevise{By this design, all three optimizations are randomly mixed to produce candidate HLO modules.} When $\beta$ is large, there is a higher probability to fuse more ops in one step. %When $\beta$ is small, the search space become larger for finding better candidate HLO modules. 
Parameter $\alpha$ determines pruning of the search space, since candidate HLO modules whose costs are larger than $\alpha$ times the cost of the best HLO module are eliminated from further exploration. Value of $\alpha$ decides a trade-off between the search time and performance of the best HLO module identified: a smaller $\alpha$ allows the search to end sooner with less recursive optimizations,  %a simple greedy algorithm without backtracking, 
 while a larger value enables exploring the search space more to potentially identify better HLO modules. We will evaluate the effects of $\alpha$ and $\beta$ in Sec.~\ref{sec:eval_alpha_beta}.

\section{Implementation}
\label{sec:implementation}

\SysName~is implemented based on JAX 0.2.3 \cite{jax2018github}. JAX is an XLA-based programming framework for generating high-performance accelerator code from pure Python and Numpy ML programs. \SysName~is implemented as a Python module that developers can readily import into their code. Core design of \SysName~is generally applicable and can be implemented in other ML frameworks as well, %. \SysName~can be integrated with other frameworks 
as a plugin module in their graph-level optimization pass.
 %\cite{chen2015mxnet}.

%\vspace{1mm}
%\noindent\textbf{Group core processing logic into a single HLO module.} 
By default, TensorFlow XLA groups the ops in a DNN model into several clusters; it generates an individual HLO module for each cluster for further optimization passes (e.g. op fusion, common sub-expression elimination and dead code elimination), separately. % In this case, the optimization for each HLO module can be conducted in parallel to save time, however, it 
This may lose the opportunity for global optimization. In single-device training, it might be still acceptable; in distributed training, joint computation and communication optimization plays an important role and has to be considered in a global view. Therefore, we build \SysName~on JAX rather than directly based on XLA, since JAX is able to group all the core processing logic into a single HLO module for further optimization.

\subsection{Activator} 
\label{sec:mpi}
The activator is implemented as a module inside XLA. %using 312 LoC of C++. 
The activator on the master node reads the optimized HLO module from a configuration file (written by the strategy maker), and sends the HLO module to all other workers using \texttt{MPIBroadcast}. %The activator in each worker waits for the HLO module from the master node. 
%After the broadcast, all workers execute the optimized HLO module together.

\vspace{1mm}
\noindent\textbf{Multi-GPU training with JAX.} Although JAX supports multi-TPU-server training, it does not support multi-machine training using GPU servers currently. To enable multi-GPU-server training with JAX, we manually modify the logic of creating the communication channels from AllReduce among GPUs on a single machine to among GPUs across multiple machines, %. The AllReduce operator is built 
based on NVIDIA Collective Communications Library (NCCL) \cite{jeaugey2017nccl}. In single-machine training, 
a unique identifer, \texttt{unique\_nccl\_id}, is created and used to create an AllReduce communication channel among multiple GPUs on the machine. In a multi-machine scenario, when the master node creates \texttt{unique\_nccl\_id}, we use \texttt{MPIBroadcast} to broadcast it to all other workers. The workers %wait until receiving \texttt{unique\_nccl\_id}, 
then create the inter-machine communication channel
based on the global \texttt{unique\_nccl\_id}. One communication channel is established for %each specific type of collective
AllReduce instructions among the same set of workers, using the same aggregation/reduce topology. %In this way, collective AllReduce can be carried out across multiple GPU servers with JAX.

\subsection{Strategy Maker} 
\label{sec:simul_imp}
%The backtracking algorithm is implemented in C++ with 564 LoC. %We set $\alpha$ to 1.05 and $\beta$ to 10 which achieve good performance as will be shown in our evaluation.
%\subsection{Simulator} 
%The simulator is built in Python with 1215 LoC (not including LoC for the components below). %The simulator simulates training process of the converted HLO module. It maintains a ready queue, consisting of instructions in execution order, whose dependencies have been cleared. It keeps removing an available instruction from the head of each ready queue, calculating completion time of the operation according to completion time of its dependencies, and adding its child nodes into the ready queue if their dependencies are all cleared. The simulator estimate the data transferring time based on the cost model built on Profiler. When an AllReduce instruction is executed, the transfer time of the AllReduce instruction can be overlapped with the execution time of other computation instructions.

\vspace{1mm}
\noindent\textbf{Profiler} is implemented based on XLA's built-in profiler by adding flag \texttt{--xla\_hlo\_profile} to the environment variable \texttt{XLA\_FLAGS}. An op may consist of multiple GPU kernels; the profiler aggregates the execution time of related kernels to obtain an accurate estimation of execution time for each op.

\vspace{1mm}
\noindent\textbf{Fused Op Estimator} is implemented in Python with 2812 LoC based on Deep Graph Library (DGL) \cite{wang2019deep}. We use 6 graph convolution layers to generate original and fused op embeddings and 3 dense layers for regression. For supervised learning of the GNN model, we randomly generate different fused ops in a number of DNN models (VGG19, ResNet50, Transformer, RNNLM, BERT and Reformer). \rerevise{We generate 30,000 samples for each DNN model. To generate a sample, we randomly select an op and fuse it with one of its predecessors, and then repeatedly fuse this fused op with one predecessor for N times, where N is randomly selected from 1,000 to 50,000. }
We train the GNN model using one Tesla V100 GPU, and it takes around 14 hours till convergence. Note that this is the time to train the base GNN model from scratch. The 6 types of DNN models contain most representative types of original ops. When predicting the execution time of a fused op that contains ops not covered in these models, we fine-tune the GNN with the new op's information, which takes much less time.

\begin{figure*}
  \centering
  \includegraphics[width=\columnwidth]{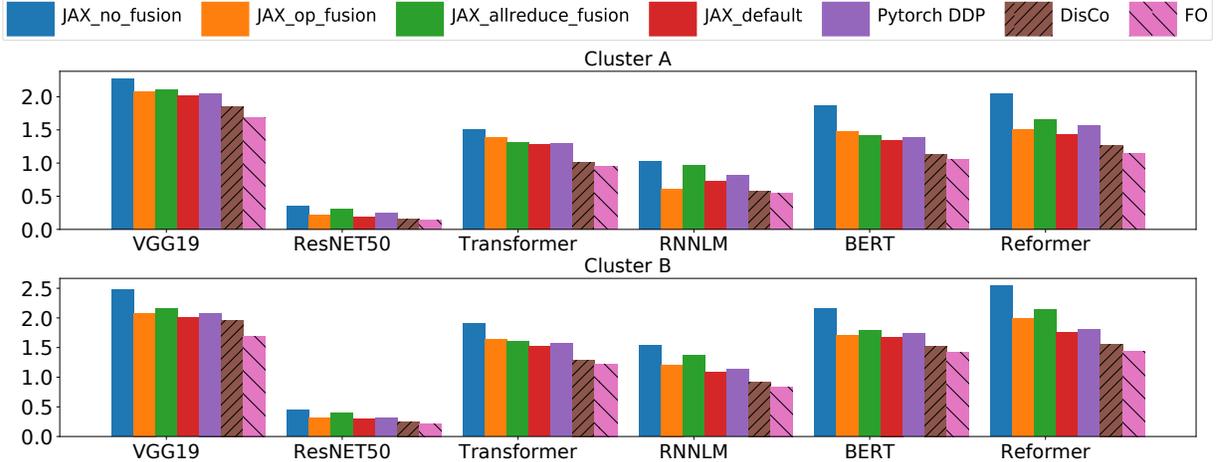}
  \vspace{-5mm}
\caption{\rerevise{Per-iteration training time comparison in Clusters A and B.}}
  \label{fig:baseline}
\end{figure*}

\section{Evaluation}

\subsection{Methodology}
\noindent\textbf{Testbed.} We evaluate \SysName~in \rerevise{2} clusters. {\bf Cluster A} consists of 6 physical machines (12 GPUs): each machine is equipped with two 11GB NVIDIA GTX 1080 Ti GPUs, one 8-core Intel Xeon E5-1660 v4 CPU and one 100GbE Mellanox RDMA card; all machines are connected through a 100GbE switch. %{\bf Cluster B} consists of 8 physical machines (32 GPUs): each machine has four 11GB NVIDIA RTX 2080 Ti GPUs, one 8-core Intel Xeon E5-1660 v4 CPU and one 10GbE NIC; all machines are connected under a 10GbE switch. 
{\bf Cluster B} consists of 8 physical machines (64 GPUs): each machine is equipped with 8 16GB NVIDIA TESLA T4 GPUs, one 96-core Intel Xeon CPU and one 100GbE NIC.

\vspace{1mm}
\noindent\textbf{Benchmark models.} We evaluate \SysName~by training 2 types of CNN models (VGG19 \cite{simonyan2014very}, ResNet \cite{he2016deep}) and 4 types of NLP models (Transformer \cite{vaswani2017attention}, RNNLM \cite{ji2016latent}, Bert \cite{devlin2018bert} and Reformer \cite{kitaev2020reformer}). Each model is trained using data parallelism with all GPUs in each cluster, based on produced strategies. %  produced strategies, we run real-world distributed training of each DNN model on our testbed according to the strategies.

%\noindent\textbf{GNN-based prediction model training.} %We profile these benchmark models, extract the fused ops in these models, generate the adjacency matrix and the feature matrix for each fused op as input to the GNN and 
%We train the GNN model using one Tesla V100 GPU, and it takes around 4 hours for the training to converge. %With the trained GNN, \SysName~can provide accurate execution time prediction of fused ops during the search process. We conduct experiments to evaluate the generality of \SysName~for unseen fused ops in Sec.~\ref{sec:general}.

\vspace{1mm}
\noindent\textbf{Baselines.} We compare \SysName~ with the following baselines. (1) {\bf JAX\_no\_fusion}: JAX with neither op nor AllReduce fusion; (2) {\bf JAX\_op\_fusion}: JAX with XLA default heuristic op fusion, which extensively fuses ops according to a post order of ops in the DNN graph, when the ops are fusible, \revise{(this baseline represents the cases of single-device op fusion optimization combined with distributed training using AllReduce).}
(3) {\bf JAX\_AllReduce\_fusion}: JAX with XLA default heuristic AllReduce fusion, which fuses neighboring AllReduce instructions based on a pre-defined tensor size threshold;
(4) {\bf JAX\_default}: JAX with XLA default heuristic op and AllReduce fusion strategies. %We modify JAX to enable multi-server GPU training (as in Sec.~\ref{sec:mpi}) in all the baselines.
\rerevise{(5) {\bf Pytorch DDP}: PyTorch DPP \cite{Li2020PyTorchD} overlaps AllReduce with the backward and forward passes. It does not consider op fusion.}

We further compare \SysName's op fusion with those in representative single-device DL compilers, TVM \cite{chen2018tvm}, nGraph \cite{cyphers2018intel} and TASO \cite{jia2019taso}.

\revise{%We do not compare \SysName{} with distributed training frameworks such as ByteScheduler \cite{peng2019generic} and Horovod \cite{sergeev2018horovod}, because the former focuses on the scheduling of communication tensors and the latter considers tensor fusion only, without op fusion optimization. 
We compare with JAX instead of XLA-enabled Tensorflow \cite{abadi2016tensorflow}, because JAX outperforms original XLA-enabled Tensorflow by grouping almost all ops into one cluster for global jit compilation optimization \cite{jax2018github}}.

\vspace{1mm}
\noindent\textbf{Default setting.} Unless stated otherwise, we use $\alpha=1.05$ and $\beta=10$ in \SysName's search algorithm. % and carry out experiments in cluster A.
To train each DNN model in a cluster, we use a batch size that can maximally exploit capacities of the respective GPU. The rationale is that if a single GPU is not fully utilized, there is no need to scale the training to many GPUs; we may just reduce the number of GPUs in use while fully utilizing each GPU.

\subsection{Training Speed-up}
In Fig.~\ref{fig:baseline}, we compare the average per-iteration training time of different models trained on our \rerevise{two} clusters, using strategies produced by \SysName~and the \rerevise{five} baselines. We observe that \SysName~always performs the best. The fully overlapping ({\bf FO}) execution time is given as a performance upper bound (i.e., lower bound of per-iteration training time), computed by maximally overlapping computation and communication without considering their inter-dependencies. Table~\ref{tab:compare} summarizes the speed-ups of \SysName, \rerevise{computed by $(T_{min}-T_{Disco})/T_{Disco}$, where $T_{min}$ is the minimum per-iteration training time achieved among the baselines and $T_{Disco}$ is the per-iteration time of \SysName.} %dividing the difference of per-iteration training time of \SysName~and the minimum time achieved among the baselines by the per-iteration time of \SysName.
It also lists the speed-ups achieved in the FO cases, computed by dividing the difference of FO's per-iteration training time and the minimum time achieved among the baselines by FO's per-iteration training time.
\revise{We do not include the search time when computing the training speed-up because the search is done offline and identified strategies can be used during the entire training process. Further, the search time is much smaller than the entire training time (e.g., within a few hours vs.~several days).}

\begin{table}[t]
    \caption{Speed-ups of \SysName~and the FO case compared to the best performance among the baselines.}
    \centering
    %\begin{small}
    \begin{tabular}{|l|l|l|l|l|l|l|l|l|l|l|}
    \hline
    \multirow{2}*{Models}&\multicolumn{2}{c|}{Cluster A}&\multicolumn{2}{c|}{Cluster B}  \\ \cline{2-5}          
    ~&\SysName&FO  &\SysName&FO        \\ \hline

    VGG19  & 8.6\% & 18.9\%& 10.1\%  &    12.5\%         \\ \hline
    ResNet50 &  12.5\%  & 28.5\%&  9.6\%& 16.8\%         \\ \hline
    Transformer  &  26.7\%  & 34.7\%&20.6\%& 25.9\%       \\ \hline
    RNNLM  &  5.1\%  & 10.9\%&8.6\%& 12.3\%           \\ \hline
    BERT  & 18.5\%  & 27.6\%& 13.7\%& 19.5\%            \\ \hline
    Reformer  &  13.4\%  &25.4\%& 14.5\%& 21.8\%            \\ \hline

    \end{tabular}%
    %\end{small}
    \label{tab:compare}
\end{table}

\begin{figure}
  \centering
  \includegraphics[width=0.75\columnwidth]{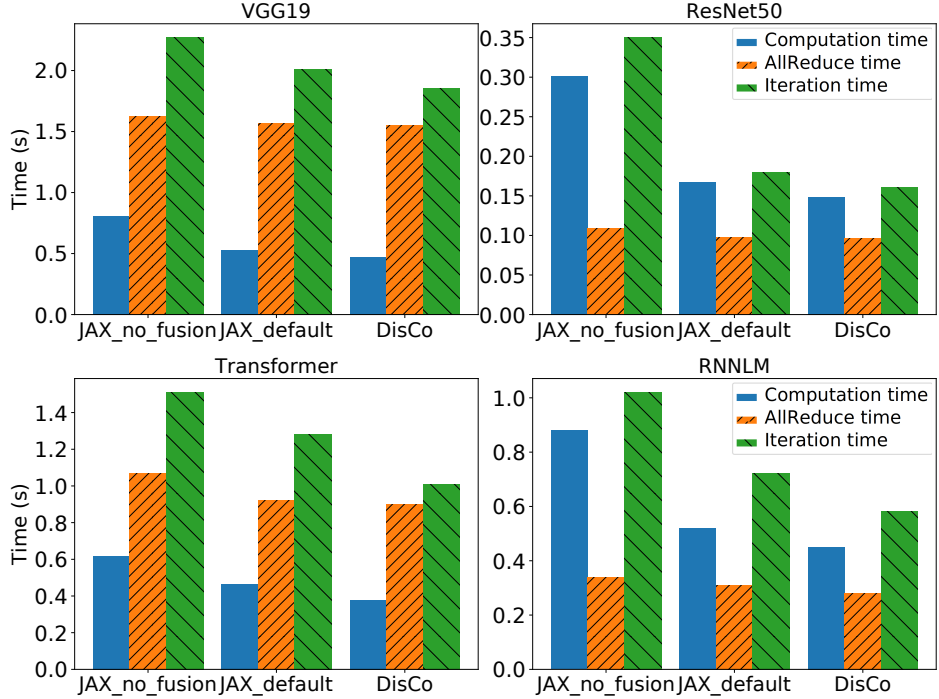}
  \caption{Per-iteration computation/communication time.} %breakdown
  \vspace{-3mm}
  \label{fig:subtime}
\end{figure}

\subsection{Time Breakdown}
Fig.~\ref{fig:subtime} shows the average per-iteration training time, computation time and communication time of training 4 models in cluster A using baselines and \SysName. Due to computation-communication overlap, per-iteration training time is usually smaller than the sum of computation time and communication time. With \SysName, computation time is smaller than that of the baselines, due to \SysName's selection of ops to fuse without a pre-defined order nor deterministic heuristics, that identifies better strategies within enlarged search space (JAX\_default adopts heuristic fusion strategies); communication time is also reduced due to our search for good AllReduce fusion strategies %to reduce the overhead by frequently transferring tiny tensors 
(JAX\_default adopts a fixed tensor size threshold to fuse AllReduce instructions). 

Regarding communication-computation overlap, for the example of Transformer, the ratio of the sum of computation and communication time over per-iteration training time is 1.12 with JAX\_no\_fusion, 1.08 with JAX\_default, and 1.27 with \SysName. These show that although JAX\_default achieves better performance than JAX\_no\_fusion in terms of computation time and communication time separately, the overlap ratio drops, because its greedy op fusion delays a large amount of communication till the completion of fused ops. %Its better performance results from the reduce computation time of the fusion rather than the overlapping between communication and computation. 
\SysName~not only reduces computation time and communication time separately, but also increases the overlap ratio by jointly choosing appropriate ops and tensors to fuse.
\SysName~achieves better performance with both computation-bound models (ResNet50 and RNNLM) and communication-bound models (VGG19 and Transformer). \rerevise{ We also notice that \SysName~usually achieves better improvement for communication-bound models than computation-bound models. It is becuase that for computation-bound models, the main benefits come from the better fusion strategy. For communication-bound models such as Transformer, the main benefits arise from the better fusion strategy, the better AllReduce fusion strategy and the better overlapping of the communication and computation. Therefore the improvement is usually more with communication-bound models.}

\begin{figure}
  \centering
  \includegraphics[width=0.5\columnwidth]{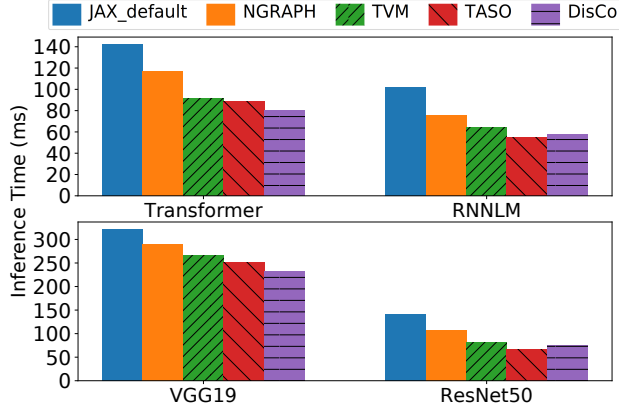}
  \caption{Comparison of single-device inference time with representative DL compilers.}
  \label{fig:single}
\end{figure}

\subsection{Single-device performance comparison}
Since most of the existing DL compilers focus on single-device training/inference acceleration, we also run \SysName~on a single device (a GTX 1080 Ti GPU) to compare the model inference time achieved with \SysName~ and with representative DL compilers. % to further evaluate \SysName's op fusion strategies. 
JAX\_default, %~\cite{jax2018github},
nGraph \cite{cyphers2018intel} and TVM \cite{chen2018tvm} use rule-based heuristics for op fusion. TASO \cite{jia2019taso} uses a search-based algorithm for graph substitution, which generates %and verifies 
subgraph candidates and then %uses a backtracking algorithm to 
searches for the best graph substitution.
Fig.~\ref{fig:single} shows that \SysName~outperforms all rule-based compilers, due to identifying better op fusion strategies using the backtracking algorithm in a larger search space, while rule-based heuristics rely on the limited number of pre-defined rules.
It achieves similar performance as TASO (slightly better with Transformer and VGG19, and slightly worse with RNNLM and ResNet50). \SysName~and TASO %both exploit search based methods for optimization, but 
focus on different search spaces: TASO is mainly for subgraph substitution and \SysName~is on op fusion. The results show that in RNNLM and ResNet50, there might be more opportunities for subgraph substitution than op fusion for TASO to achieve better performance.

\begin{figure}
  \centering
  \includegraphics[width=0.7\linewidth]{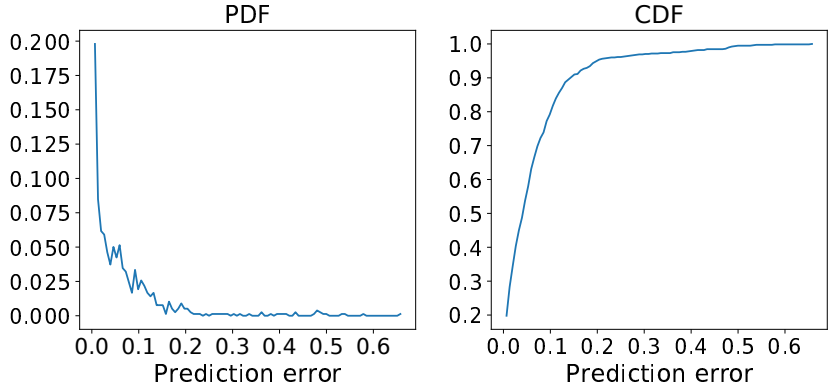}
  \caption{Probability Density Function and Cumulative Distribution Function of prediction errors of Fused Op Estimator. 
  }

  \label{fig:general}
\end{figure}

\subsection{Simulator Accuracy}%Generalization to unseen fused operators}
\label{sec:general}

We evaluate the accuracy of our GNN-based {\em Fused Op Estimator} by randomly generating 2000 unseen fused ops, which do not appear in our GNN training sample set. In this experiment, both the GNN training set and the above test samples are profiled on a GTX 1080 Ti GPU. The execution time of these fused ops ranges from 20 microseconds to 30 milliseconds. We compare the predicted execution time of fused ops in the test set %by the {\em Fused op Estimator}
and their profiled execution time, and compute a prediction error by dividing the absolute difference of these two values by the profiled execution time. Fig.~\ref{fig:general} shows the PDF and CDF of the prediction errors. We see that more than 90\% predictions are within 14\% error of the respective real execution time. It shows that the GNN-based estimator can effectively learn the structural information of fused ops, by considering data and control dependencies among the original ops in the fused op subgraphs.

\begin{table}[t]
\caption{Estimation error of the simulator.}
\centering
%\begin{small}
\begin{tabular}{|l|l|l|l|l|}
\hline
\multirow{2}*{Models}&Real Execution  &Simulation &Error  \\ 
~&Time (s) &Time (s)&  \\ \hline

VGG19        & 1.85  &    2.08      &   12.4\%        \\ \hline
ResNet50     & 0.16    &    0.18    & 11.1\%       \\ \hline
Transformer  & 1.01    &   1.15     &  13.9\%      \\ \hline
RNNLM        &  0.58   &   0.66     &  13.8\%      \\ \hline
BERT         & 1.13     &  1.3      &  15.9\%     \\ \hline
Reformer     &  1.26    &    1.48   &  17.5\%      \\ \hline
\end{tabular}
%\end{small}
\label{tab:simulator}
\end{table}

We then test the accuracy of the simulator for estimating the end-to-end execution time of HLO modules. % based on the {\em Fused Op Estimator}. 
Table~\ref{tab:simulator} gives the estimated time by the simulator (simulation time) to execute the best HLO module found by \SysName~ on each DNN model and the respective real execution time in cluster A. The error is calculated by dividing the absolute difference of simulation time and real execution time by the real execution time. The simulator achieves a 11.1\% error ratio for RNNLM and at most 17.5\% for Reformer, which is good enough to guide the search algorithm. %in the {\em Strategy Maker}. 
It also implies that the linear regression model for estimating the execution time of AllReduce instructions is accurate enough in spite of its simple form.

\subsection{Effects of optimization methods}

\begin{figure}
  \centering
  \includegraphics[width=0.5\columnwidth]{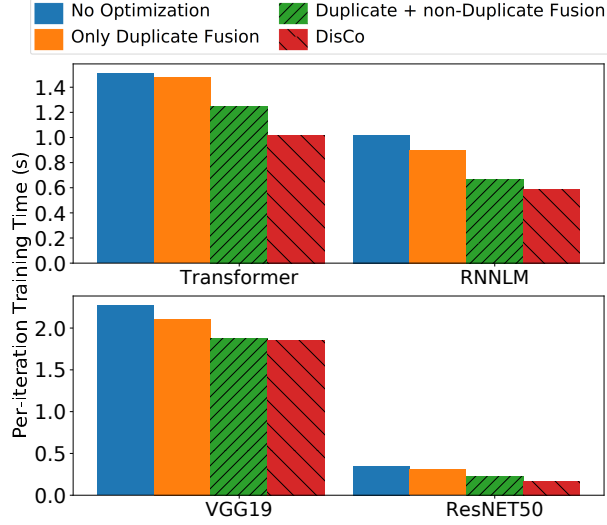}
  \caption{\rerevise{\SysName~%performance
  with/without certain optimizations.}}
  \label{fig:submodule}
\end{figure}

We further evaluate the design of three optimization methods in \SysName~(Sec.~\ref{sec:searchalg}), duplicate op fusion, non-duplicate op fusion and AllReduce fusion, by \rerevise{increasingly adding} each one of them in the search algorithm, respectively, %We disable one optimization dimension from the duplicate op fusion, non-duplicate op fusion, AllReduce fusion in \SysName~
when training %Transformer, RNNLM, VGG19 and ResNet50 
DNNs in cluster A. Fig.~\ref{fig:submodule} shows that each optimization positively contributes to training time reduction,  %compared with the baseline without any optimization and none of them can achieve the performance as good as \SysName,
and the joint application of all three achieves the best performance.  We further observe that non-duplicate fusion plays the most important role in reducing the training time, since per-iteration time %increases
\rerevise{decreases} most when it is %disabled
\rerevise{added}, as compared to %disabling
\rerevise{adding} either of the other two methods. %, when training each of the DNN models.
This is because compared with non-duplicate fusion, duplicate fusion leads to extra computation, and is usually suitable when the output of the predecessor op needs to be sent to other successors earlier (especially when the successor is an AllReduce instruction); in most cases, the output of a predecessor op is activation that does not need to be transferred to other devices. We also notice that in case of VGG19, \SysName's performance is similar with and without AllReduce fusion. This is because most of the large gradient tensors in VGG19 are from the fully connected layers and are transferred at the beginning of back propagation; after transferring these large tensors, communication of other small tensors can overlap well with computation (with op fusion that our search algorithm identifies) even without tensor fusion.

\subsection{Parameters in backtracking algorithm}
\label{sec:eval_alpha_beta}

\begin{table}[!t]
\caption{Per-iteration time and strategy search time with different values of $\alpha$}
\centering
\begin{small}
\begin{tabular}{|l|l|l|l|l|}
\hline
\multirow{2}*{Models}&\multicolumn{3}{c|}{Execution Time(s)/Search Time(min)} \\ \cline{2-4}
&$\alpha=1$ &$\alpha=1.05$ &$\alpha=1.1$  \\ \hline
VGG19        & 2.01/11       &   1.85/17         &    1.83/36           \\ \hline
ResNet50     & 0.18/15       &  0.16/22          &    0.16/48    \\ \hline
Transformer  & 1.18/54      &   1.01/74         &   1.02/143      \\ \hline
RNNLM        &  0.74/6      &   0.58/9         &   0.57/16      \\ \hline
BERT         & 1.26/69      &    1.13/89        &  1.10/158       \\ \hline
Reformer     &  1.44/53     &   1.26/78        &    1.21/161       \\ \hline

\end{tabular}
\end{small}
\label{tab:search}
\end{table}

We tune parameters $\alpha$ and $\beta$ in our search algorithm (Alg.~\ref{alg:search}), and train DNNs in cluster A with the respective best strategies. % found by the search algorithm.
Table~\ref{tab:search} and Table~\ref{tab:search1} show the result per-iteration training time, along with the search time to find the respective best strategy on each DNN model.

Setting $\beta$ to 10 %which we found performs well in most of experiments, 
and varying $\alpha$, Table~\ref{tab:search} shows that with a larger $\alpha$, training time decreases (because the search space is larger with more candidate HLO modules enqueued and repeatedly optimized), while the search time increases accordingly. $\alpha=1.05$ leads to a good trade-off between strategy quality and search time.

\begin{table}[t]
\caption{Per-iteration training time and strategy search time with different values of $\beta$}
\centering
%\begin{small}
\begin{tabular}{|l|l|l|l|l|}
\hline
\multirow{2}*{Models}&\multicolumn{4}{c|}{Execution Time(s)/Search Time(min)} \\ \cline{2-5}
~&$\beta=1$ &$\beta=5$ &$\beta=10$&$\beta=30$   \\ \hline
VGG19        & 1.83/66     &    1.87/29    &   1.85/17       &    2.09/14        \\ \hline
ResNet50     & 0.12/84     &    0.15/41    &  0.16/22        &     0.21/15     \\ \hline
Transformer  & 1.00/195     &   1.02/98     &   1.01/74       &   1.14/54   \\ \hline
RNNLM        &  0.53/64    &   0.54/17     &   0.58/9       &   0.76/7     \\ \hline
BERT         & 1.12/258     &  1.14/114      &    1.13/89      &  1.26/71     \\ \hline
Reformer     &  1.19/239    &    1.24/101    &   1.26/78       &    1.38/59     \\ \hline

\end{tabular}
%\end{small}

\label{tab:search1}
\end{table}

Setting $\alpha$ to 1.05 and varying $\beta$, Table~\ref{tab:search1} shows that when $\beta$ increases, the search time decreases (because there is a higher probability to fuse more ops within one algorithm step, reducing the search space), while the training time increases in general. %the trade-off is that the best found per-iteration training time of these models usually decreases with the increase of the value of $\beta$. 
We observed that when $\beta$ is relatively smaller (i.e. ranging from 1 to 10), the training time increases slowly but the search time drops significantly with the increase of $\beta$. When $\beta$ is small, the modification of the HLO module is subtle in each step, leading to slow search progress in a huge search space. Training time increases with larger $\beta$ because in each step, there is a higher probability for \SysName~to carry out op fusion for multiple times before execution time of the produced HLO module is evaluated, which may miss part of the search space to find a better strategy. We identify $\beta=10$ to be a good choice for the trade-off between training performance and search time.

\section{Related Work}
\subsection{Deep Learning Compiler}

%\noindent \textbf{Deep Learning Compilers.} Hummingbird \cite{Hummingbird} compiles featurization process and traditional ML models (e.g., decision trees) into a small groups of operations. 
%Rammer \cite{RAMMER} generates an efficient static spatio-temporal schedule for a DNN at compile time to minimize scheduling overhead. %It maximizes hardware utilization by exploiting parallelism on both inter- and intra- operator co-scheduling.
MLIR~\cite{lattner2020mlir} is a reusable and extensible compiler infrastructure that standardizes the Static Single Assignment-based IR data structures and provides a declarative system to define IR dialects. Relay \cite{roesch2019relay} presents a compiler framework to unify and generalize IR in existing frameworks. %expressing state-of-the-art models by its functional, statically typed IR. 
Intel nGraph \cite{cyphers2018intel} simplifies the realization of optimized DL performance across software frameworks and hardware platforms with carefully designed IR and bridge to connect different frameworks. It carries out op fusion extensively similar to XLA's approach.
TVM \cite{chen2018tvm} is a compiler that exposes graph-level and operator-level optimizations to provide performance portability for DL workloads across diverse hardware backends.  
It defines four types of ops (injective, reduction, complex-out-fusible, and opaque), and provides generic rules to fuse these ops: multiple injective ops can be fused into another injective op; a reduction op can be fused with input injective ops; ops such as conv2d are complex-out-fusible, and their outputs can be fused with element-wise ops.  
All these compilers %mainly 
focus on DNN training/inference on a single device. %Instead, \SysName~accelerates DNN training in distributed environment. 

For distributed DNN compilation, XLA integrates collective AllReduce into its HLO module; the AllReduce optimization pass and op fusion optimization pass adopt simple heuristics and are done separately. % without considering the interactions between them.
Gshard \cite{lepikhin2020gshard} is an extension of XLA which provides convenient APIs for sharding large models; no extra op fusion and AllReduce fusion strategies are provided. Boehm {\em et al.}~\cite{boehm2018optimizing} provide distributed op primitives in their customized compiler, but focus on traditional ML jobs, e.g., training KMeans \cite{krishna1999genetic} or L2SVM \cite{mu2007breast}. % rather than DL model training. They %use tree search to decide op fusion candidates by 
They divide a model graph into several parts and use tree search to decide the fusion strategy separately for each part; %within a smaller strategy space; 
this may lose the opportunity for global optimization.

\begin{comment}
\subsection{Operator fusion}

Most of compilers apply rule based heuristics for operation fusion. TVM \cite{chen2018tvm} defines four types of operators (i.e. injective 2. reduction, 3. complex-out-fusible, and opaque), and provide generic rules to fuse these operators: multiple injective operators can be fused into another injective operator; a reduction operator can be fused with input injective operators; operators such as conv2d are complex-out-fusible, and their outputs can be fused with element-wise operators. These rules are applied to transform the computational graph into a fused version. For example, Intel nGraph \cite{cyphers2018intel} and MXNet \cite{chen2015mxnet} also design some preliminary heuristics for operator fusion, e.g., $a \times b + 1$ is replaced by a single BLAS or GPU call. Fusionstitching \cite{long2018fusionstitching} fuse operators with critical-path reduction based heuristics, and considers both producer-consumer fusion opportunities and also fusion of fine-grained ops in the same DNN layer. Google team \cite{abdolrashidi2019learning} proposed a learning-based framework to decide the order of operator fusion. However, Most of these fusion strategies are only applicable in single device scenario rather than distributed clusters without considering the potential communication delay brought by their fusion strategies.  
\end{comment}

\subsection{Learning-based prediction model}
%\begin{comment}
Several learning-based cost models have been developed for automatic code optimization. MILEPOST GCC (GNU Compiler Collection) \cite{fursin2011milepost} uses a 1-nearest-neighbor model which takes as input manually selected features and predicts the best combinations of compiler flags for GCC. Ithemal \cite{mendis2019ithemal} uses an LSTM model to predict the throughput of assembly-level code. Baghdadi {\em et al.}~\cite{learningbased} integrate a DL-based cost model %for automatic code optimization, which is integrated 
into an auto-scheduler, that enables the Tiramisu compiler to select the best code transformation for a given program. 
%\end{comment}

In the area of DNN optimization, Kaufman {\em et al.}~\cite{kaufman2020learned} introduce a GNN-based method for a number of optimization decisions (e.g., tile-size selection, operator fusion), %performance models 
based on tensor computation graphs for TPU-based training. %They use GNN to directly encode operation properties and generates embedding vector for each node in the graph. The vector is further used for different optimization tasks such as tile-size selection and operator fusion. 
DynaTune \cite{zhangdynatune} designs a Bayesian belief model to predict the potential performance gain of each operator with uncertainty quantification, to guide the optimization process of finding better fusion strategies. %Although so many learning based cost models are proposed, 
 We are the first in integrating a GNN-based fused op cost model for joint op and tenor fusion optimization.

\subsection{Distributed Neural Network Training}

Horovod \cite{sergeev2018horovod} decouples communication from specific training frameworks and optimizes it using tensor fusion. A tensor fusion threshold \texttt{HOROVOD\_FUSION\_THRESHOLD} is pre-defined, and small AllReduce tensors are combined within this size threshold for transmission, which is similar to XLA's tensor fusion approach. %However, the decoupling makes it hard to jointly optimize op fusion and communication together. 
%Parallax \cite{kim2019parallax} advocates different gradient aggregation methods for different types of parameters in models with large word embeddings: PS for sparse parameters and AllReduce for dense parameters. 
ByteScheduler \cite{peng2019generic} advocates a priority-based tensor scheduling strategy for better communication-computation overlapping; no interaction with computation op fusion is considered. \rererevise{GShard \cite{lepikhin2020gshard} creates shards of weights and model states that can be split among ranks. CoCoNet \cite{jangda2022breaking} introduces a domain-specific language to easily express communication and computation in distributed training. Megatron-LM \cite{shoeybi2019megatron} introduces an efficient intra-layer model-parallel approach to support training of very large transformer models. GPipe \cite{huang2018gpipe} uses pipelining to address memory bottlenecks for training large NNs. PipeDream \cite{harlap2018pipedream} introduces a pipelining design to overlap communication and computation for asynchronous training with convergence guarantee. These projects are focusing on model or pipeline parallelism which is orthogonal to \SysName{}.}

\section{Concluding Discussions}
\label{sec:con}

We present \SysName, a deep learning compiler based on JAX for distributed DNN training acceleration. % which is totally transparent for model developers without changing one line of code in their models. 
\SysName~jointly optimizes computation operator fusion and AllReduce tensor fusion using a backtracking search algorithm, maximizing the overlap of computation and communication and minimizing overall training time. A GNN-based simulator is built to effectively facilitate the search in large joint op/tensor fusion strategy space. \SysName~achieves good training speed-up as compared with existing fusion schemes and the full communication-computation overlap case, in typical distributed environments.

As a future direction, we plan to extend \SysName~from data-parallel training to supporting model parallelism and pipeline parallelism. %With data parallelism, \SysName~only need to profile the DNN model on a single device and estimates per-iteration training time on single-device because the replica models on each devices are the same.
To accelerate DNN model training using model or pipeline parallelism with joint op and tensor fusion, we first need to improve our simulator: we shall profile model training on all devices %rather than on only one replica in data parallelism. The simulator also needs to
and measure activation transfer time across devices as well. The optimization methods in the search algorithm should also be expanded to include fusion of activations. The HLO module will include send and recv communication instructions for activations, besides AllReduce instructions for gradient tensors. Further, the design of \SysName{} can be readily extended to handle the parameter server architecture for tensor communication, by replacing AllReduce instructions with push and pull communication instructions, while the dependencies between push and pull are readily included in the HLO module.

%Bibliography
\bibliographystyle{unsrt}  
\bibliography{main}

\end{document}